\documentstyle[epsfig]{elsart}
\begin{document}
\begin{frontmatter}
\hfill
$\mbox{\small{\begin{tabular}{r}
${\rm Napoli-DSF-99-11}$
\end{tabular}}}$

\vspace{1cm}

%_______________________ Title, Authors ____________________________________
%{\hspace*{\fill}\parbox{55mm}
\title{Leptonic and Semileptonic Decays of Pseudoscalar Mesons}
\author[bltp]{M.A. Ivanov}
\author{and}
\author[infn]{P. Santorelli}
\address[bltp]{Bogoliubov Laboratory of Theoretical Physics, \\
Joint Institute for Nuclear Research, 141980 Dubna, Russia}
\address[infn]{Dipartimento di Scienze Fisiche, \\
Universit\`a  ``Federico II" di Napoli, Napoli, Italy\\
and\\
INFN Sezione di Napoli}
%-------------------------------------------------------------------
\begin{abstract}
We employ the relativistic constituent quark model to give a unified
description of the leptonic and semileptonic decays of pseudoscalar
mesons ($\pi$, $K$, $D$, $D_s$, $B$, $B_s$).
The calculated leptonic decay constants and form factors are found to be
in good agreement with available experimental data and other approaches.
We reproduce the results of spin-flavor symmetry in the heavy quark
limit.
\end{abstract}
%\begin{keyword}
%Pseudoscalar mesons, Leptonic and semileptonic decays;
%Quark models; Spin-flavor symmetry.\\
%{\sc PACS}: 13.20.-v, 13.20.He, 12.38.Lg, 24.85.+p
%\end{keyword}
\end{frontmatter}
%____________________________________________________________

\section{Introduction}

Semileptonic decays of pseudoscalar mesons allow to evaluate the
elements of the Cabibbo-Kobayashi-Maskawa (CKM) matrix, which are
fundamental parameters of the Standard Model. The decay $K\to\pi e\nu$
provides the most accurate determination of $V_{us}$, the semileptonic
decays of D and B mesons, $D\to K(K^*) l\nu$, $B\to D(D^*)l\nu$ and
$B\to \pi(\rho)l\nu$, can be used to determine $|V_{cs}|$, $|V_{cb}|$
and $|V_{ub}|$, respectively. The effects of strong interactions in
these processes can be expressed in terms of form factors, which depend
on $q^2$, the squared momentum transferred to the leptonic pair.
Information on the form factors are obtained by measuring the
distributions of $q^2$ and decay angles.

The decays of heavy D and B mesons are of particular interest due to the
spin-flavor symmetry observed for infinite quark masses \cite{IW}. This
symmetry allows to reduce the number of form factors and express them in
terms of the universal Isgur-Wise function \cite{IWf}. Also the scaling
laws derived for some physical observables can be, in principle, tested
experimentally. Since the Isgur-Wise function cannot be calculated from
first principles, many models and nonperturbative approaches, which
exhibit the heavy quark symmetry, have been employed to describe
relevant phenomena. However, it was found out, that the finite mass
corrections are very important, especially, in the charm sector. It
appears that in some sense a step back should be done from using
the heavy quark symmetry as a guide under model building to the 
straightforward calculations with full quark propagators. 
Then one has to check the consistency of the
results with the spin-flavor symmetry in the heavy quark limit.

In this paper we employ the relativistic constituent quark model (RCQM)
\cite{RCQM}  for the simultaneous description of both light and heavy
flavored meson leptonic and semileptonic decays. 
This model is based
on the effective Lagrangian describing the coupling of mesons with their
quark constituents, and the compositeness condition.
The physical processes are described by the one-loop quark diagrams
with free constituent propagators and meson-quark vertices
related to the Bethe-Salpeter amplitudes. The masses of lower-lying
pseudoscalar (PS) mesons should be  less than the sum of quark
constituent masses to provide the absence of imaginary parts
corresponding to quark production. The adjustable parameters, the widths
of Bethe-Salpeter amplitudes in momentum space  and constituent quark
masses, are determined from the best fit of available experimental data
and some lattice determinations. We found that our results are in good
agreement with experimental data and other approaches. Also we
reproduce the results of spin-flavor symmetry for leptonic decay
constants and semileptonic form factors in the heavy quark limit.

The shapes of vertex functions and quark propagators should be found
from the Bethe-Salpeter and Dyson-Schwinger equations, respectively.
This is provided by the Dyson-Schwinger Equation (DSE) \cite{DSE}
studies. A DSE-approach has been employed to provide a unified and
uniformly accurate description of light- and heavy-meson observables
\cite{DSEH1,DSEH2}.

A similar approach, based on the effective heavy meson Lagrangian, has been
described in \cite{Gatto} in terms of a model based on meson-quark
interactions, where mesonic  transition amplitudes are represented by
diagrams with heavy mesons attached to quark loops.
The free propagator has been used for light quarks.
However, the quark propagator obtained in the heavy quark limit has been
employed for heavy quarks.

\section{The model}

We employ an approach \cite{RCQM} based on the effective interaction
Lagrangian which describes the transition of hadron into quarks. For
example, the transition of the meson $H$ into its constituents $q_1$ and
$q_2$ is given by the Lagrangian

\begin{equation}
\label{lag}
{\cal L}_{{\rm int}} (x)=g_H H(x) \int\!\! dx_1 \!\!\int\!\! dx_2
\Phi_H (x;x_1,x_2)
\bar q(x_1) \Gamma_H \lambda_H q(x_2)\,.
\end{equation}
Here, $\lambda_H$ and $\Gamma_H$ are the Gell-Mann and Dirac matrices,
respectively, which provide the flavor and spin numbers of the meson
$H$. The function $\Phi_H$ is related to the scalar part of
Bethe-Salpeter amplitude. For instance, the separable form
$\Phi_H(x;x_1,x_2)=\delta(x-(x_1+x_2)/2) f((x_1-x_2)^2)$ has been used
in \cite{RCQM} for pions.

The coupling constants $g_H$ is given by the so called {\it
compositeness condition} proposed in \cite{SW} and extensively used in
\cite{EI}. That condition  means that the renormalization constant of
the meson field is equal to zero:
\begin{equation}
\label{comp}
Z_H=1-\frac{3g^2_H}{4\pi^2}\tilde\Pi^\prime_H(m^2_H)=0\,,
\end{equation}
where $\tilde\Pi^\prime_H$ is the derivative of the meson mass operator
defined by

\begin{equation}
\label{mass}
\tilde\Pi_H(p^2)=\int\!\!{d^4k\over 4\pi^2i}\phi_H^2(-k^2)
{\rm tr}\biggl[\Gamma_H S_2(\not\! k) \Gamma_H S_1(\not\! k+\not\! p)\biggr].
\end{equation}

The invariant amplitudes describing the leptonic $H(p)\to l\nu$ and
semileptonic $H(p)\to H'(p^\prime) l\nu$ decays are written down

\begin{eqnarray}
A(H(p) \to e \nu)&=&
{ G_F \over \sqrt{2} }
V_{qq'}
(\bar e O_{\mu}\nu) M_H^\mu(p)
\label{matlep}\\
&&\nonumber\\
A(H(p)\to H'(p') e\nu)&=&{G_F\over \sqrt{2}}V_{qq'}(\bar e O_{\mu}\nu)
M^\mu_{HH'}(p,p^\prime),
\label{matsem}
\end{eqnarray}
where $G_F$ is the Fermi weak-decay constant, $V_{qq'}$ is the appropriate
element of the CKM matrix.
The matrix elements of the hadronic currents are given by

\begin{eqnarray}
M_H^\mu(p)&=&
{3\over 4\pi^2}g_H\int\!\!{d^4k\over 4\pi^2i}\phi_H(-k^2)
{\rm tr}\biggl[\gamma^5 S_2(\not\! k)O^\mu S_1(\not\! k+\not\! p)\biggr]=
f_H p^\mu
\label{curlep}\\
&&\nonumber\\
M^{\mu}_{HH'}(p,p^\prime)&=&
{3\over 4\pi^2}g_Hg_{H'}\!\!\int\!\!{d^4k\over 4\pi^2i}
\phi_H(-k^2)\phi_{H'}(-k^2)
\label{cursem}\\
&&\times
{\rm tr}\biggl[\gamma^5 S_3(\not\! k)\gamma^5
S_2(\not\! k+\not\! p^\prime)O^\mu
S_1(\not\! k+\not\! p) \biggr]
\nonumber\\
&&\nonumber\\
&&=f_+(q^2)(p+p^\prime)^\mu + f_-(q^2)(p-p^\prime)^\mu\\
\nonumber
\end{eqnarray}
where $\phi_H(-k^2)$ is related to the BS-amplitude in momentum space, and

\begin{equation}\label{prop}
S_i(\not\! k)=\frac{1}{m_i-\not\! k}
\end{equation}
is the propagator of the constituent quark with mass $m_i$. As discussed
before, to avoid the appearance of imaginary parts in
Eqs.~(\ref{curlep}) and (\ref{cursem}), we assume that
$m_H<m_{q_1}+m_{q_2}$ which is a reliable approximation for the
lower-lying mesons considered here.

To evaluate the integral in Eq.~(\ref{cursem})

\begin{equation}\label{int}
I_{HH'}(p,p^\prime)=\int \frac{d^4k}{4\pi^2i}
{\cal F}(-k^2)
{\rm tr}
\biggl
\{\gamma^5 S_3(\not\! k)\gamma^5
S_2(\not\! k+\not\! p^\prime)\gamma^\mu S_1(\not\! k+\not\! p)
\biggr\}\,,
\end{equation}
where
${\cal F}(-k^2)=\phi_H(-k^2)\cdot\phi_{H^\prime}(-k^2)$,
we need to calculate the following integrals:

\begin{equation}\label{int1}
J^{(0,\mu,\mu\nu,\mu\nu\delta)}=\int \frac{d^4k}{\pi^2i}
\frac{(1,k^\mu,k^\mu k^\nu,k^\mu k^\nu k^\delta){\cal F}(-k^2)}
{[m_1^2-(k+p)^2][m_2^2-(k+p^\prime)^2][m_3^2-k^2]}\,.
\end{equation}
Using the Cauchy representation for the function ${\cal F}(-k^2)$ and then
the standard techniques of the Feynman $\alpha-$parametrization one finds
(${\cal F'}(z)=d{\cal F}(z)/dz$)

\begin{eqnarray}
J^{0}&=&
\int\limits_0^\infty dt \biggl(\frac{t}{1+t}\biggr)^2
\int\! d^3\alpha\,\delta\biggl(1-\sum\limits_{i=1}^3\alpha_i\biggr)
\biggl(-{\cal F'}(z_I)\biggr)
\label{scalar}\\
&&\nonumber\\
&&\nonumber\\
J^{\mu}&=&
-\int\limits_0^\infty dt \biggl(\frac{t}{1+t}\biggr)^3
\int\! d^3\alpha\,\delta\biggl(1-\sum\limits_{i=1}^3\alpha_i\biggr)
P_\alpha^\mu
\biggl(-{\cal F'}(z_I)\biggr)
\label{vector}\\
&&\nonumber\\
&&\nonumber\\
J^{\mu\nu}&=&
\int\limits_0^\infty dt \biggl(\frac{t}{1+t}\biggr)^2
\int\! d^3\alpha\, \delta\biggl(1-\sum\limits_{i=1}^3\alpha_i\biggr)
\label{tensor2}\\
&&\nonumber\\
&&\times\biggl\{
-\frac{1}{2}g^{\mu\nu}\frac{1}{1+t}{\cal F}(z_I)
-P_\alpha^\mu P_\alpha^\nu \biggl(\frac{t}{1+t}\biggr)^2
{\cal F'}(z_I)
\biggr\}
\nonumber\\
&&\nonumber\\
&&\nonumber\\
J^{\mu\nu\delta}&=&
\int\limits_0^\infty dt \biggl(\frac{t}{1+t}\biggr)^2
\int\! d^3\alpha\, \delta\biggl(1-\sum\limits_{i=1}^3\alpha_i\biggr)
\label{tensor3}\\
&&\nonumber\\
&&\times\biggl\{
\frac{1}{2}\biggl[g^{\mu\nu}P_\alpha^\delta+
                  g^{\mu\delta}P_\alpha^\nu+
                  g^{\nu\delta}P_\alpha^\mu \biggr]
\frac{t}{(1+t)^2}{\cal F}(z_I)
\nonumber\\
&&\nonumber\\
&&+P_\alpha^\mu P_\alpha^\nu P_\alpha^\delta \biggl(\frac{t}{1+t}\biggr)^3
{\cal F'}(z_I)
\biggr\}
\nonumber
\end{eqnarray}
where $q=p-p^\prime$,  $P_\alpha=\alpha_1 p+\alpha_2 p^\prime$,
$D_3=\alpha_1\alpha_3 p^2+\alpha_2\alpha_3 p^{\prime 2}+\alpha_1\alpha_2 q^2 $,
and $z_I=t[\sum_{i=1}^3\alpha_i m^2_i-D_3]-P_\alpha^2t/(1+t)$.

Finally, Eq.~(\ref{int}) becomes
$$ %\begin{equation}
I_{HH'}^\mu(p,p^\prime)=
(p+p^\prime)^\mu\, I_+(p^2,p^{\prime 2},q^2)+
(p-p^\prime)^\mu\, I_-(p^2,p^{\prime 2},q^2)
$$ %\end{equation}
with
\begin{eqnarray}
I_+(p^2,p^{\prime 2},q^2)&=&
\frac{1}{2}\int\limits_0^\infty dt \biggl(\frac{t}{1+t}\biggr)^2
\int\! d^3\alpha\, \delta\biggl(1-\sum\limits_{i=1}^3\alpha_i\biggr)
\label{intfin}\\
&&\nonumber\\
&\times&
\biggr\{
{\cal F}(z_I)\frac{1}{1+t}\biggl[4-3(\alpha_1+\alpha_2)\frac{t}{1+t}\biggr]
\nonumber\\
&&\nonumber\\
&&
-{\cal F'}(z_I) \biggl[(m_1+m_2)m_3
\nonumber\\
&&\nonumber\\
&&
+\frac{t}{1+t}\biggl(-(\alpha_1+\alpha_2)(m_1m_3+m_2m_3-m_1m_2)
\nonumber\\
&&\nonumber\\
&&
+\alpha_1 p^2+\alpha_2 p^{\prime 2}\biggr)
-P_\alpha^2 \biggl(\frac{t}{1+t}\biggr)^2
\biggl(2-(\alpha_1+\alpha_2)\frac{t}{1+t}\biggr)
\biggr]
\biggr\}\,.
\nonumber
\end{eqnarray}
The normalization condition is written in the form

\begin{equation} \label{normfin}
\frac{3g_H^2}{4\pi^2}I_+(p^2,p^2,0)=1
\end{equation}
with $m_1=m_2\equiv m$.

The integrals corresponding to the matrix element of the leptonic
decay $H(p)\to l\nu$ and radiative decay of neutral meson
$H(p)\to\gamma(q_1)+\gamma(q_2)$ are calculated following the same
procedure. We have

\begin{eqnarray}
Y^\mu(p)&=&
\int \frac{d^4k}{4\pi^2i}\phi(-k^2)
{\rm tr}\biggl\{\gamma^5 S_2(\not\! k)\gamma^\mu(I-\gamma^5)
S_1(\not\! k+\not\! p)\biggr\}=p^\mu Y(p^2)
\nonumber\\
&&\nonumber\\
Y(p^2)&=&
\int\limits_0^\infty dt \frac{t}{(1+t)^2} \int\limits_0^1 d\alpha
\biggl[m_2+(m_1-m_2)\frac{\alpha t}{1+t}\biggr]\phi(z_Y)
\label{intlep}\\
&&\nonumber\\
&&\nonumber\\
K^{\mu\nu}(q_1,q_2)&=&
\int \frac{d^4k}{4\pi^2i}
\phi(-k^2)
{\rm tr}
\biggl\{\gamma^5 S(\not\! k-\not\! q_2)\gamma^\mu
S(\not\! k) \gamma^\nu S(\not\! k+\not\! q_1)\biggr\}
\nonumber\\
&&\nonumber\\
&=&
i\varepsilon^{\mu\nu\alpha\beta}q_1^\alpha q_2^\beta  K(p^2)
\nonumber\\
&&\nonumber\\
K(p^2)&=&
m\int\limits_0^\infty dt \biggl(\frac{t}{1+t}\biggr)^2
\int\limits_0^1 d\alpha_1 \int\limits_0^{1-\alpha_1} d\alpha_2
\biggl(-\phi'(z_K)\biggr)
\label{intrad}
\end{eqnarray}
where
$z_Y=t[\alpha m_1^2+(1-\alpha)m_2^2-\alpha p^2+\alpha^2 p^2t /(1+t)]$
and
$z_K=t[ m_1^2-\alpha_1\alpha_2 p^2]+\alpha_1\alpha_2 p^2 t/(1+t)$.

The physical observables are expressed in terms of the structural integrals
written in Eqs.~(\ref{intfin}), (\ref{intlep}) and (\ref{intrad}):
\begin{eqnarray}
g_{P\gamma\gamma}=\frac{g_P}{2\sqrt{2}\pi^2}K(m^2_P), &&
\hspace{1cm}
\Gamma(P\to \gamma\gamma)=\frac{\pi}{4}\alpha^2 m_P^3 g_{P\gamma\gamma}^2,
\label{rad}\\
&&\nonumber\\
&&\nonumber\\
f_P = \frac{3}{4\pi^2}\ g_P\ Y(m^2_P), &&
\Gamma(P\to  l\nu)=|V_{qq'}|^2\frac{G_F^2 f_P^2}{8\pi} m_P m_l^2
\biggl[1-\frac{m_l^2}{m_P^2}\biggr]^2,
\label{lep}\\
&&\nonumber\\
&&\nonumber\\
f_+(q^2)&=&\frac{3}{4\pi^2}\ g_Pg_{P'}\ I_+(m^2_P,m^2_{P'},q^2),
\label{sem}\\
&&\nonumber\\
\Gamma(P\to P^\prime l\nu)&=&|V_{qq'}|^2\frac{G_F^2}{192\pi^3 m_P^3}
\int\limits_0^{t_-} dt |f_+(t)|^2 \biggl[(t_+-t)(t_--t)\biggr]^{3/2},
\nonumber
\end{eqnarray}
with $t_\pm=(m_P\pm m_{P^\prime})^2$ (the extra factor 1/2 appears for
$\pi^0$ in the final state).

\subsection{Heavy quark limit}

The leptonic heavy decay constants and semileptonic heavy to heavy form
factors acquire a simple form in the heavy quark limit, {\it i. e.} when
$m_1\equiv M\to\infty$, $m_2\equiv M'\to\infty$ and $p^2=(M+E)^2$,
$p^{\prime 2}=(M'+E)^2$ with $E$ being a constant value. From
Eq.~(\ref{intfin}) by replacing the variables $\alpha_1\to\alpha_1/M$
and $\alpha_2\to\alpha_2/M'$, one obtains

\begin{eqnarray}
I_+ & \rightarrow & \frac{M+M'}{2MM'}\cdot
\int\limits_0^\infty dt \biggl(\frac{t}{1+t}\biggr)^2
\int\limits_0^1 d\alpha\alpha \int\limits_0^1 d\tau
\biggl({-\cal F'}(z)\biggr)
\biggl[m+\frac{\alpha t}{1+t}\biggr]
\nonumber\\
&&\nonumber\\
&=& \frac{M+M'}{2MM'}\cdot \frac{1}{2}
\int\limits_0^1 \frac{d\tau}{W}
\int\limits_0^\infty du
{\cal F}(\tilde z)\frac{m+\sqrt{u}}{m^2+\tilde z}
\label{inthql}
\end{eqnarray}
where $\tilde z=u-2E\sqrt{u/W}$,  $W=1+2\tau(1-\tau)(w-1)$ and
$w=(M^2+M^{\prime 2}-2MM' q^2)/(2MM')$.

The normalization condition can be obtained from Eq.~(\ref{inthql}) by putting
$w=1$ and $M'=M$. We have

\begin{equation}\label{normhql}
\frac{3g_H^2}{4\pi^2}\cdot I_+^{(0)}=1,
\hspace{0.8cm}
I_+^{(0)}=\frac{1}{2M}I_N,
\hspace{0.8cm}
I_N= \int\limits_0^\infty du \phi_H^2(\tilde z_0)
\frac{m+\sqrt{u}}{m^2+\tilde z_0}
\end{equation}
where $\tilde z_0=u-2E\sqrt{u}$.
Then the leptonic decay constant and semileptonic form factors are written as

\begin{eqnarray}
f_P & \rightarrow & \frac{1}{\sqrt{M}}\cdot \sqrt{\frac{3}{2\pi^2 I_N}}
\int\limits_0^\infty du [\sqrt{u}-E]\phi_H(\tilde z_0)
\frac{m+\sqrt{u}/2}{m^2+\tilde z_0}
\label{lephql}\\
&&\nonumber\\
f_{\pm}& \rightarrow & \frac{M'\pm M}{2\sqrt{MM'}}\xi(w)
\hspace{1cm}
\xi(w)=\frac{1}{I_N}\int\limits_0^1 \frac{d\tau}{W}
\int\limits_0^\infty du \phi_H^2(\tilde z)
\frac{m+\sqrt{u}}{m^2+\tilde z}\,.
\label{semhql}
\end{eqnarray}
It is readily seen that we reproduce the scaling law for both leptonic
decay constants and form factors, and obtain the explicit expression
for the Isgur-Wise function \cite{IW,IWf}.

\section{Results and discussion}

The expressions obtained in the previous section for physical
observables are valid for any vertex function $\phi_H(-k^2)$. Here, we
choose a Gaussian form $\phi(-k^2)=\exp\{k^2/\Lambda_H^2\}$ in Minkowski
space. The magnitude  of $\Lambda_H$ characterizes the size of the
BS-amplitude and  is an adjustable parameter in our approach. Thus, we
have six $\Lambda$-parameters plus the four quark masses, all of which
are fixed via the least-squares fit to the observables measured
experimentally or taken from lattice simulations (see,
Table~\ref{t1}).

\begin{table}[t]
\caption{Calculated values of a range of observables
($g_{\pi\gamma\gamma}$ in GeV$^{-1}$, leptonic decay constants in GeV,
form factors and ratios are dimensionless). The ``Obs." are extracted from
Refs.~\protect \cite{PDG,CLEO-BD,CLEO-Bpi,Flynn,Wittig,Debbio,MILC}.
The quantities used in fitting our parameters are marked by ``$\ast$".
\label{t1}}
\begin{tabular}{clll|clll}
  &      & Obs.  & Calc. & & & Obs.  & Calc. \\
\hline
$\ast$ & $g_{\pi\gamma\gamma}$   & 0.274                  & 0.242   &
       & $f_+^{K\pi}(0)$         & 0.98                   & 0.98    \\
$\ast$ & $f_\pi$                 & 0.131                  & 0.131   &
$\ast$ & $f_+^{DK}(0)$           & 0.74 $\pm$ 0.03        & 0.74    \\
$\ast$ & $f_K$                   & 0.160                  & 0.160   &
       & $f_+^{BD}(0)$           &                        & 0.73    \\
$\ast$ & $f_D$                   & 0.191$^{+19}_{-28}$    & 0.191   &
       & $f_+^{B\pi}(0)$         & 0.27 $\pm$ 0.11        & 0.51    \\
$\ast$ & $\frac{f_{D_s}}{f_D}$   & 1.08(8)                & 1.08    &
  & ${\rm Br}(K\to\pi l\nu)$&   $(4.82\pm 0.06)\cdot 10^{-2}$
                            &   $4.4\cdot 10^{-2}$            \\
  & $f_{D_s}$               & 0.206$^{+18}_{-28}$     & 0.206   &
  & ${\rm Br}(D\to K l\nu)$ &   $(6.8\pm 0.8)\cdot 10^{-2}$
                            &   $ 8.1\cdot 10^{-2}$            \\
$\ast$ & $f_B$                   & 0.172$^{+27}_{-31}$     & 0.172   &
       & ${\rm Br}(B\to D l\nu)$ &   $(2.00\pm 0.25)\cdot 10^{-2}$
                            &   $2.3\cdot 10^{-2}$            \\
$\ast$ & $\frac{f_{B_s}}{f_B}$   & 1.14(8)                & 1.14    &
       & ${\rm Br}(B\to\pi l\nu)$&   $(1.8\pm 0.6)\cdot 10^{-4}$
                            &   $2.1\cdot 10^{-4}$            \\
  & $f_{B_s}$               &                        & 0.196   &
  &                         &                        &         \\
\hline
\end{tabular}
\end{table}

The fit yields the values of $\Lambda$-parameters and the constituent
quark masses which are listed in Eqs.~(\ref{fitlam}) and (\ref{fitmas}).

\begin{equation}\label{fitlam}
\begin{array}{ccccccc}
\Lambda_\pi & \Lambda_K & \Lambda_D & \Lambda_{D_s} &
\Lambda_B & \Lambda_{B_s} & {\rm (in\ GeV)}\\
\hline
$\ \ 1.16\ \ $ & $\ \ 1.82\ \ $ & $\ \ 1.87\ \ $ &
$\ \ 1.95\ \ $ & $\ \ 2.16\ \ $ & $\ \ 2.27\ \ $ & \\
\end{array}
\end{equation}

\begin{equation}\label{fitmas}
\begin{array}{ccccc}
m_u & m_s & m_c & m_b & {\rm (in\ GeV)} \\
\hline
$\ \ 0.235\ \ $ & $\ \ 0.333\ \ $ &  $\ \ 1.67\ \ $ & $\ \ 5.06\ \ $ &  \\
\end{array}
\end{equation}

The values  of $\Lambda$ are such that $\Lambda_{m_i}<\Lambda_{m_j}$ if
$m_i<m_j$. This corresponds to the ordering law for sizes of bound
states. The values of $\Lambda_D=1.87$ GeV and $\Lambda_B=2.16$ GeV are
larger than those obtained in \cite{DSEH2}: $\Lambda_D=1.41$ GeV and
$\Lambda_B=1.65$ GeV. The mass of u-quark and the parameter
$\Lambda_\pi$ are almost fixed from the decays $\pi\to\mu\nu$ and
$\pi^0\to\gamma\gamma$ with an accuracy of a few percent. The obtained
value of the u-quark mass $m_u=0.235$ GeV is less than the
constituent-light-quark mass typically employed in quark models for
baryon physics ($m_u>m_N/3=0.313$ GeV). For instance, the value of
$m_u=0.420$ GeV was extracted from fitting nucleon observables within
our approach \cite{RCQM}. The different choice of constituent quark
masses is a common feature of quark models with free propagators due to
the lack of confinement. However, we consider here the low-lying mesons
that allows us to fix the constituent quark masses in a self-consistent
manner. As mentioned above, the meson masses must be less than the sum
of masses of their constituents. This gives the restrictions on the choice
of the meson binding energies: $E_K=m_K-m_s<m_u$, $E_D=m_D-m_c<m_u$ and
$E_B=m_B-m_b<m_u$, which means that the binding energy cannot be
relatively large as compared with those obtained in \cite{DSEH2}:
$E_D=0.58$ GeV and $E_B=0.74$ GeV.

Let us now consider the $q^2$-behaviour of the form factors.
We use the three-parameter function for the four $f_+$ form factors
\begin{equation}\label{approx}
f_+^{HH'}(q^2)=\frac {f(0)} {1-b_0(q^2/m_H^2)-b_1(q^2/m_H^2)^2}
\end{equation}
here $b_0~,b_1~$ and $f(0)$ are parameters to be fitted. We collect the
fitted values in the following Table and report the $q^2$-dependence in
Fig. 1.

\begin{equation}
\label{coef}
\begin{array}{c|cccc}
            &\ \ K\to \pi\ \  & \ \ D\to K\ \  &\ \ B\to D\ \
&\ \ B\to\pi\ \  \\
\hline
     b_0    & 0.28  & 0.64 & 0.77 & 0.52 \\
     b_1    & 0.057 & 0.20 & 0.19 & 0.38 \\
\end{array}\,
\end{equation}

For comparison, we plot, together with our results,
the predictions of a vector dominance monopole model:

\begin{equation}\label{mon}
f_+^{q\to q'}(q^2)=\frac{f_+^{q\to q'}(0)} {1-q^2/m^2_{V_{qq'}}}
\end{equation}
with $m^2_{V_{qq'}}$ being a mass of lower-lying $\bar qq'$-vector meson.
We choose
$m_{D^*_s}=2.11$ GeV  for $c\to s$,
$m_{B^*}=5.325$ GeV   for $b\to u$,
$m_{B_c^*}\approx m_{B_c}=6.4$ GeV \cite{CDF-Bc}  for $b\to c$ transitions.
The values of $f_+^{qq'}(0)$ are taken from the Table 1. Also we calculate
the  branching ratios of semileptonic decays  by using widely accepted values
of the CKM matrix  elements \cite{PDG}.

Our result for the slope of the $K_{l3}$ form factor 

\begin{equation}
\lambda_+ = m_{\pi}^2 \frac{f_+^{K\pi\prime}(0)}{f_+^{K\pi}(0)} = 0.023\ ,
\end{equation}
is in good agreement with experiment:
$\lambda_+^{\rm expt} =  0.0286 \pm 0.0022$
\cite{PDG} and VDM prediction:
$\lambda_+^{\rm VDM} = m_{\pi}^2/m^2_{K^{\ast}} = 0.025.$
This value is also consistent with Refs. \cite{kl3}

One can see that the agreement with experimental data and lattice
results is very good, with the exception of the value of $f_+^{bu}(0)$
which is found to be larger  than the monopole extrapolation of a
lattice simulation, QCD Sum Rules (cf. \cite{CSB}) and some other quark
models (see, for example, \cite{BSW,LNS}). However, this result is
consistent with the value calculated from Refs.~\cite{DSEH2,infra} and
allows us to reproduce the experimental data for $B\to \pi l\nu$ with
quite good accuracy. 

%\vspace{2cm}
%
%{\large\bf Acknowledgments}

\section*{Acknowledgments}

We appreciate F. Buccella, V.E. Lyubovitskij, G. Nardulli and C.D.
Roberts for many interesting discussions and critical remarks. We thank
G. Esposito for reading the manuscript. M.A.I. gratefully acknowledges
the hospitality and support of the Theory Group at Naples University
where this work was conducted. This work was supported in part by the
Russian Fund for Fundamental Research, under contract number
99-02-17731-a.

%______________________________ References ______________________________
%

\newpage

\begin{figure}[t]
\begin{center}
\begin{tabular}{cc}
\epsfig{file=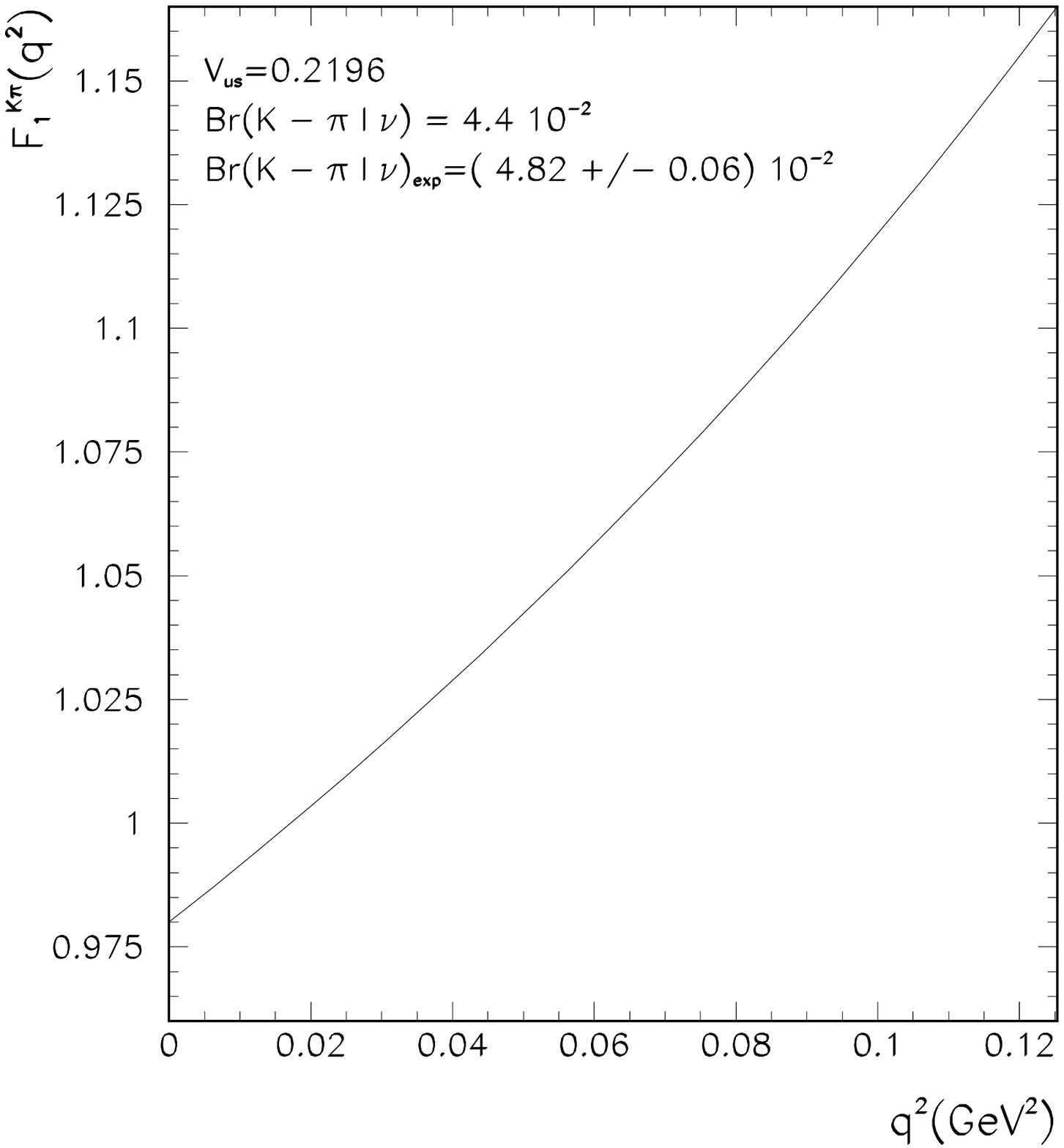,height=8cm} &
\epsfig{file=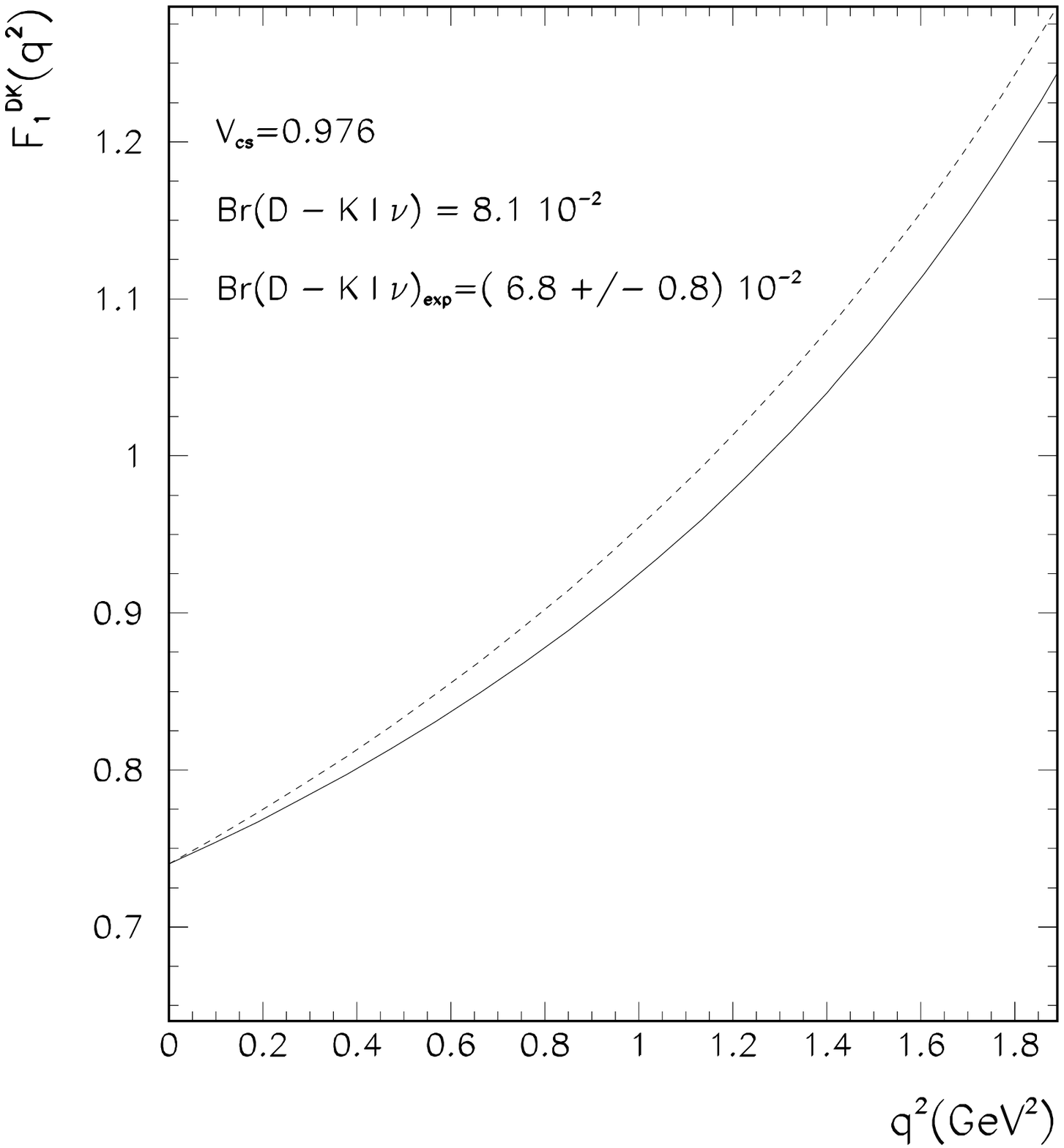,height=8cm} \\
\epsfig{file=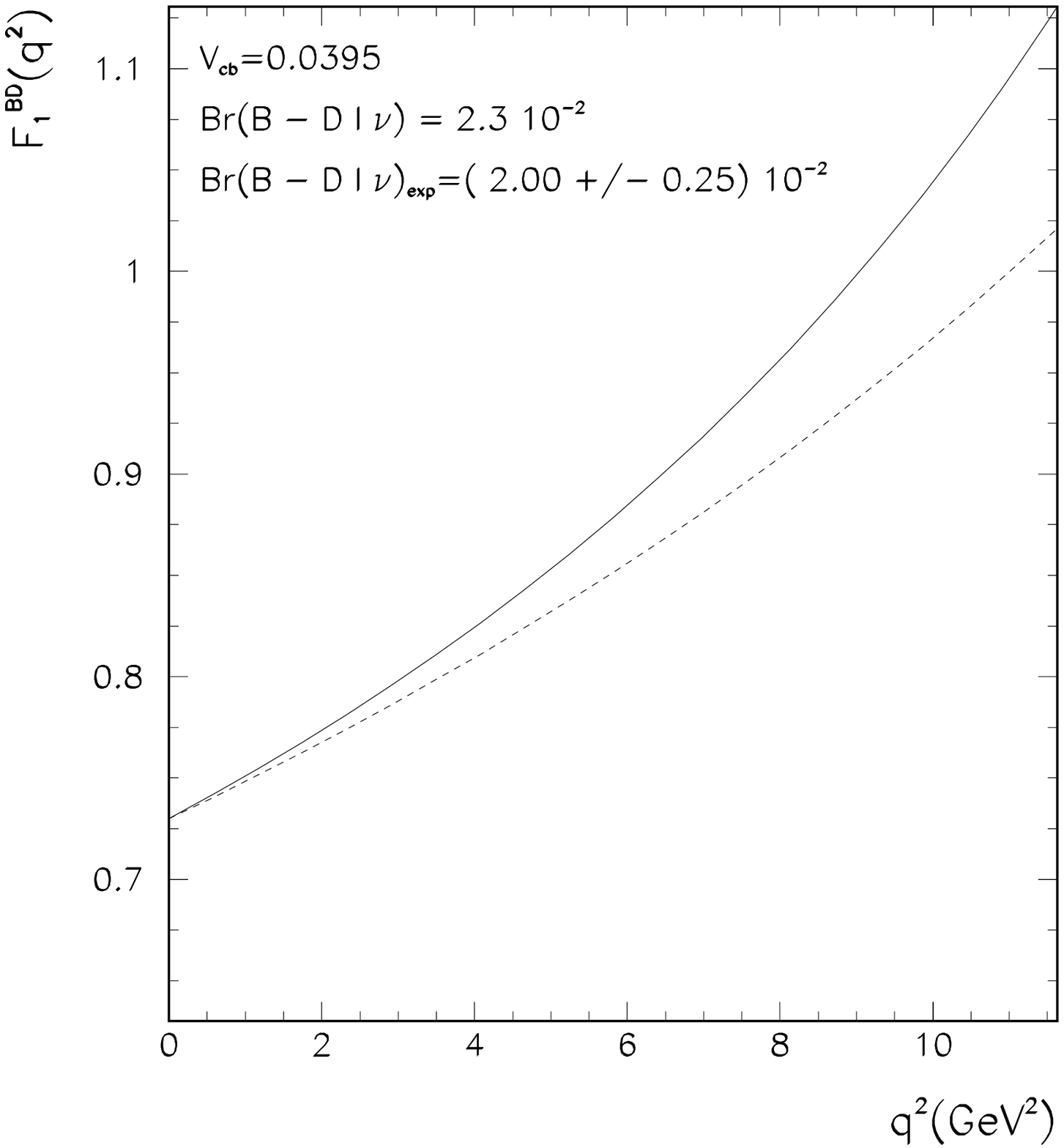,height=8cm} &
\epsfig{file=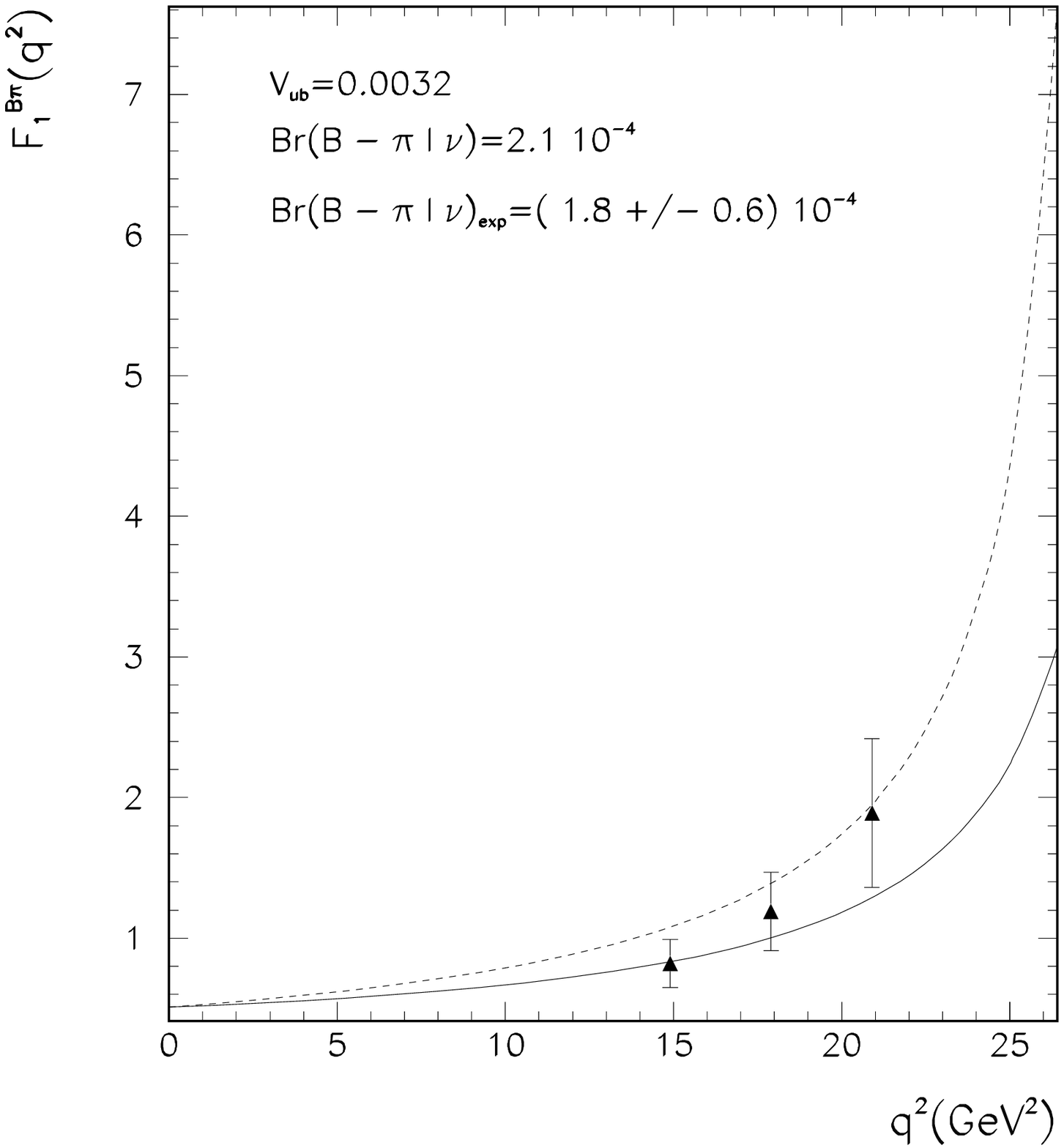,height=8cm}
\end{tabular}
\end{center}
\caption
{
The semileptonic $K\to\pi$, $D\to K$, $B\to D$ and $B\to\pi$
form factors with, for comparison, a vector dominance, monopole model
Eq.~(\ref{mon}) and a lattice simulation \protect\cite{Burford}.
Our results: continuous lines. Monopole: dotted lines. Lattice: data points.
}
\label{f:fig1}
\end{figure}

\end{document}